\documentclass[preprint,12pt,authoryear]{elsarticle}
\usepackage{url} 
\usepackage{lineno}
\usepackage{color}  

\usepackage{amsmath}
\usepackage{amssymb}
\usepackage{scalerel}
\usepackage{setspace}
\usepackage{pifont}
\usepackage{fontawesome}
\usepackage{xcolor}
\usepackage{soul}
\usepackage{placeins}
\usepackage{verbatim}
\usepackage{todonotes}
\usepackage{newtxmath}
\journal{ }
\begin{document}

\begin{frontmatter}

\title{Physics-Informed Convolutional Decoder (PICD): A novel approach for direct inversion of heterogeneous subsurface flow}

\author[inst1,inst2]{Nanzhe Wang}
\author[inst2]{Xiang-Zhao Kong\corref{cor1}}
\author[inst3,inst4]{Dongxiao Zhang\corref{cor2}}

\cortext[cor1]{Corresponding author: xkong@ethz.ch (Xiang-Zhao Kong)}
\cortext[cor2]{Corresponding author: zhangdx@sustech.edu.cn (Dongxiao Zhang)}

\affiliation[inst1]{organization={Department of Energy Science and Engineering, Stanford University},
            city={Palo Alto},
            postcode={CA 94305}, 
            country={U.S.A.}}

\affiliation[inst2]{organization={Geothermal Energy and Geofluids Group, Institute of Geophysics, ETH Zurich},
            addressline={Sonneggstrasse 5}, 
            city={Zurich},
            postcode={8092}, 
            country={Switzerland}}

\affiliation[inst3]{organization={Eastern Institute for Advanced Study, Eastern Institute of Technology},
            city={Ningbo, Zhejiang},
            postcode={315200}, 
            country={P. R. China}}
          
\affiliation[inst4]{organization={National Center for Applied Mathematics Shenzhen (NCAMS), Southern University of Science and Technology},
            city={Shenzhen, Guangdong},
            postcode={518000}, 
            country={P. R. China}}

\begin{abstract} 
In this study, we present the development and application of the physics-informed convolutional decoder (PICD) framework for inverse modeling of heterogenous groundwater flow. PICD stands out as a direct inversion method, eliminating the need for repeated forward model simulations. The framework leverages both data-driven and physics-driven approaches by integrating monitoring data and domain knowledge (governing equation, boundary conditions, and initial conditions) into the inversion process. PICD utilizes a convolutional decoder to effectively approximate the spatial distribution of hydraulic heads, while Karhunen–Loeve expansion (KLE) is employed to parameterize hydraulic conductivities. During the training process, the stochastic vector in KLE and the parameters of the convolutional decoder are adjusted simultaneously, ensuring that the predictions align with available measurements and adhere to domain-specific knowledge. The final optimized stochastic vectors correspond to the estimation of hydraulic conductivities, and the trained convolutional decoder demonstrates the ability to predict the evolution and distribution of hydraulic heads in heterogeneous fields. To validate the effectiveness of the proposed PICD framework, various scenarios of groundwater flow are examined. Results demonstrate the framework's capability to accurately estimate heterogeneous hydraulic conductivities and to deliver satisfactory predictions of hydraulic heads, even with sparse measurements. The proposed PICD framework emerges as a promising tool for efficient and informed groundwater flow inverse modeling.
\end{abstract}


\end{frontmatter}

\section{Introduction}
\label{sec:intro}
Groundwater flow modeling plays a crucial role in examining the subsurface movement of water, offering valuable insights for effective water resource utilization and management. However, inherent challenges arise from our limited knowledge of subsurface formations, resulting in groundwater flow models influenced by uncertain parameters, such as hydraulic conductivity. The presence of these uncertainties can introduce variations in model predictions. To address this, inverse modeling becomes imperative, enabling the estimation of uncertain parameters through the utilization of sparse measurements. The incorporation of inverse modeling not only enhances our comprehension of the underground formation properties but also facilitates improved groundwater flow modeling, contributing to more accurate predictions and informed decision-making in water resource management. 

In recent decades, inversion methods have undergone extensive scrutiny, with discussions on these techniques available in seminal works, such as \citet{mclaughlin1996reassessment}, \citet{tarantola2005inverse}, \citet{oliver2008inverse}, \citet{wang2021deep}, and \citet{wang2023deep}. Traditional inversion methods are typically categorized into two main groups: gradient-based methods and non-gradient-based methods. Gradient-based methods, including well-established techniques, such as the steepest descent method \citep{chavent1975history}, Gauss-Newton method \citep{reynolds1996reparameterization,li2003history}, Quasi-Newton method \citep{yang1988automatic,gao2006improved}, conjugate gradient method \citep{lee1986history,lee1987estimation},and Levenberg-Marquardt method \citep{arenas2001semi,zhang2003initial}, are widely employed for inversion tasks. While gradient-based methods exhibit relatively fast convergence, their drawback lies in the necessity for gradient information computation, which can be computationally expensive or impractical in certain scenarios. 

Non-gradient-based methods offer the advantage of circumventing the direct computation of gradient information, making them more adaptable to integration with numerical simulators. Within this category, intelligent optimization algorithms stand out as heuristic methods that draw inspiration from biological evolution processes or natural physical phenomena. These algorithms have been broadly used in solving inverse problems, such as Genetic Algorithm (GA) \citep{sayyafzadeh2012regularization,xavier2013genetic}, Simulated Annealing (SA) \citep{ouenes1992enhancing,jeong2019theoretical}, and Particle Swarm Optimization (PSO) \citep{mohamed2010application,lee2019field}. Markov chain Monte Carlo (MCMC) methods constitute another subset of non-gradient-based methods, offering the posterior distribution of the model parameters through sampling from the desired probability distribution. MCMC has been found applications in inverse modeling scenarios \citep{bonet1996hybrid,oliver1997markov}. However, its drawback lies in its demand for a substantial number of samplings to converge, leading to a significant computational burden. Ensemble-based methods are another class of non-gradient-based methods, which have garnered attention for inverse modeling in recent years, with the Ensemble Kalman Filter (EnKF) as a prominent example. Initially proposed by \citet{evensen1994sequential}, EnKF has been successfully applied in diverse fields, such as hydrology \citep{chen2006data,xue2014multimodel}, meteorology \citep{houtekamer2001sequential,houtekamer2005atmospheric}, and petroleum engineering \citep{gu2004history,chang2010data}. EnKF dynamically updates parameters with the acquisition of time-series measurements, allowing direct parameter updates using new measurements without the need to reevaluate all previous measurements. The Ensemble Smoother (ES) represents another widely adopted ensemble-based method for inverse problems \citep{chen2013levenberg,le2016adaptive,skjervheim2011ensemble}. Unlike the sequential update of EnKF, ES globally updates model parameters using measurements from all time-steps simultaneously. However, it is noteworthy that ensemble-based methods necessitate the simultaneous evaluation and updating of a group of parameters, entailing numerous forward modeling using simulators. 

Conventional inverse modeling methods, including gradient-based and non-gradient-based approaches, typically require iterative forward modeling to access the likelihood of the estimated unknown parameters, resulting in time-consuming and computationally demanding procedures. An alternative to this iterative process is the adoption of the ‘direct inversion method’ \citep{wang2021deep}, in which both measurements and physical laws are leveraged to directly adjust the unknown parameters. With the evolution of machine learning methods, direct inversion can be realized by incorporating domain knowledge into the machine learning models. One noteworthy development in this realm is the physics-informed neural network (PINN), an emerging method gaining prominence in recent years. PINN is capable of estimating constant parameters in partial differential equations (PDEs) by utilizing both measurements and domain knowledge \citep{raissi2019physics}. In PINN, the residuals of the governing equations are computed using automatic differentiation (AD) \citep{baydin2018automatic}, which are then minimized along with the data mismatch during the training process. In addition to constant parameters, PINN has also been utilized for estimation of space-dependent parameters, such as hydraulic conductivity \citep{tartakovsky2020physics, he2020physics}. However, it is worth noting that PINN often requires a relatively large number of measurements to achieve satisfactory performance. 

In this work, we introduce a novel machine learning (ML)-based direct inversion method, termed the Physics-Informed Convolutional Decoder (PICD) framework. While sharing the integration of measurements and domain knowledge (e.g., governing equation, boundary conditions, and initial conditions) with PINN, PICD diverges in several key aspects. Firstly, it employs a convolutional decoder network to approximate the states of hydraulic heads, instead of a fully-connected neural network. This choice is motivated by the convolutional network's enhanced capability to capture spatial information effectively. Secondly, PICD leverages Karhunen–Loeve expansion (KLE) to represent space-dependent hydraulic conductivity, which allows the incorporation of space-correlated statistical information. This, in turn, reduces the dimensionality of parameters requiring estimation. Thirdly, the physical equations of domain knowledge are discretized using the finite difference method to calculate the discretized residuals. These residuals are minimized in the training process to enforce constraints. To evaluate the performance of the proposed PICD framework, several scenarios of groundwater flow problems with diverse settings are presented. Results show that the PICD framework accurately captures and estimates the general distribution of hydraulic conductivity even with sparse measurements. In addition, the inversion with PICD yields satisfactory predictions of hydraulic head evolution.

The remainder of this paper is organized as follows. In Section~\ref{sec:Problem statement}, we introduce the governing equation of the considered groundwater flow problem and formulate the definitions of forward and inverse modeling. Section~\ref{sec:method} provides a detailed description of KLE and the proposed PICD framework. In Section~\ref{sec:results}, we assess the performance of the proposed PICD framework with diverse numerical experiments of groundwater flow. Finally, Section~\ref{sec:discussions and conclusions} presents discussions and conclusions.

\section{Problem statement}
\label{sec:Problem statement}
In this work, we focus on utilizing the proposed PICD framework in the context of groundwater flow problems. This section introduces the governing equation and the associated inverse problems. 

The following governing equation for dynamic groundwater flow is considered:
\begin{center}
\begin{equation}
{{S}_{s}}\frac{\partial h(\mathbf{x},t)}{\partial t}=\nabla \cdot \left( K(\mathbf{x})\nabla h(\mathbf{x},t) \right)+q(\mathbf{x},t),\text{     }\mathbf{x}\in D,\text{ }t\in [0,{{L}_{t}}],
\label{eq:ge}
\end{equation}
\end{center}
where ${{S}_{s}}$ [1/L] denotes the specific storage, $h(\mathbf{x},t)$ [L] denotes the hydraulic head, $K(\mathbf{x})$ [L/T] denotes the hydraulic conductivity, $q(\mathbf{x},t)$ [1/T] denotes the source or sink term, $D$ denotes the physical domain of the studied problem, and ${{L}_{t}}$ [T] denotes the total time span of the studied problem. The groundwater flow problems are additionally subjected to boundary and initial conditions, which are formulated as follow:
\begin{center}
\begin{equation}
h(\mathbf{x},t)=H(\mathbf{x}),\text{       }\mathbf{x}\in {{\Gamma }_{D}},\text{ }t\in [0,{{L}_{t}}],
\label{eq:dboun}
\end{equation}
\end{center}
\begin{center}
\begin{equation}
K(\mathbf{x})\nabla h(\mathbf{x},t)\cdot \mathbf{n}(\mathbf{x})=-Q(\mathbf{x}),\text{       }\mathbf{x}\in {{\Gamma }_{N}},\text{ }t\in [0,{{L}_{t}}],
\label{eq:nboun}
\end{equation}
\end{center}
\begin{center}
\begin{equation}
h(\mathbf{x},t=0)={{H}_{0}}(\mathbf{x}),\text{       }\mathbf{x}\in D,
\label{eq:ic}
\end{equation}
\end{center}
where $H(\mathbf{x})$ [L] denotes the specified hydraulic head on the Dirichlet boundary, ${{\Gamma }_{D}}$, of the domain, $Q(\mathbf{x})$ [L/T] denotes the specified flux on the Neumann boundary, ${{\Gamma }_{N}}$, of the domain, $\mathbf{n}(\mathbf{x})$ [-] denotes the outward unit vector perpendicular to the boundary, and ${{H}_{0}}(\mathbf{x})$ [L] denotes the initial hydraulic heads.

Given a specific distribution of hydraulic conductivity, $K(\mathbf{x})$, the partial differential equation Eq.~(\ref{eq:ge}) can be numerically solved to derive the corresponding distribution and evolution of hydraulic heads, $h(\mathbf{x},t)$, a process commonly referred to as \emph{forward modeling}. Various numerical solution methods, such as finite difference method, finite element method, and finite volume method, can be employed for this purpose.  

In practical groundwater flow problems, obtaining the specific distribution of the space-dependent hydraulic conductivity, $K(\mathbf{x})$, is usually challenging due to the inherent heterogeneity and the sparsity of available measurements. This challenge introduces significant uncertainty into the solving process of Eq.~(\ref{eq:ge}). Therefore, inferring the unknown hydraulic conductivity field, $K(\mathbf{x})$, becomes crucial to address this limitation. In real-world groundwater flow scenarios, hydraulic heads and hydraulic conductivity measurements of several sparse observation locations can be collocated. These collocated measurements offer an opportunity to estimate the overall distribution of hydraulic conductivity fields, consequently reducing the uncertainty associated with hydraulic head predictions. This process is known as \emph{inverse modeling}, and constitutes the primary focus of this work. 

In many inversion methods, a large number of forward calculations are usually required to compare the predictions with measurements and adjust the estimations of hydraulic conductivity based on the observed mismatch. However, our focus in this work is on the direct inversion method, in which the need for multiple/repetitive forward calculations is eliminated during the inversion process. 

\section{Method}
\label{sec:method}
This section introduces the proposed direct inversion method, the Physics-Informed Convolutional Decoder (PICD) framework. The PICD framework combines a convolutional decoder network and a parameterization technique. This innovative approach leverages both available measurements and domain knowledge to directly invert unknown hydraulic parameters (i.e., conductivity fields) in groundwater flow problems.

\subsection{KLE-based parameterization}
\label{sec:kle}
Recognizing the challenges posed by space-dependent and high-dimensional parameters in the estimation process, it is beneficial to employ parameterization techniques for parameter-dimension reduction. By transforming the high-dimensional hydraulic conductivity fields into low-dimensional vectors through parameterization, the complexity of the inversion task can be significantly reduced. 

In this context, we treat the hydraulic conductivity field, $K(\mathbf{x})$, as a spatially correlated stochastic field. To effectively parameterize and represent $K(\mathbf{x})$, we utilize the Karhunen–Loeve expansion (KLE) \citep{zhang2004efficient}. Assuming a log-normal distribution for $K(\mathbf{x})$, the log-transformed hydraulic conductivity, $\ln K(\mathbf{x})$, is decomposed using KLE as follows:
\begin{center}
\begin{equation}
\ln K(\mathbf{x})=\left\langle \ln K(\mathbf{x}) \right\rangle +\sum\limits_{i=1}^{\infty }{\sqrt{{{\lambda }_{i}}}{{f}_{i}}(\mathbf{x}){{\xi }_{i}}},
\label{eq:kle}
\end{equation}
\end{center}
where $\left\langle \ln K(\mathbf{x}) \right\rangle$ denotes the mean of $\ln K(\mathbf{x})$, ${{\lambda }_{i}}$ denotes the eigenvalue of the covariance of $\ln K(\mathbf{x})$, ${{f}_{i}}(\mathbf{x})$ denotes the eigenfunction of the covariance of $\ln K(\mathbf{x})$, and ${{\xi }_{i}}$ denotes independent and orthogonal Gaussian stochastic variables with zero mean and unit variance in the case where $\ln K(\mathbf{x})$ is a second-order stationary Gaussian stochastic field. The infinite expansion in the Eq.~(\ref{eq:kle}) can be truncated into finite terms. The total number of retained terms, $N$, in the expansion can be determined according to the decay rate of ${{\lambda }_{i}}$, while preserving a certain percentage (${\sum\nolimits_{i=1}^{N}{{{\lambda }_{i}}}}/{\sum\nolimits_{i=1}^{\infty }{{{\lambda }_{i}}}}\;$) of stochastic field information. Here, $\sum\nolimits_{i=1}^{\infty }{{{\lambda }_{i}}}=D\sigma _{\ln K}^{2}$ \citep{li2007probabilistic}, where $\sigma _{\ln K}^{2}$ denotes the variance of the stochastic field and $D$ denotes the domain size. Therefore, the truncated expansion of $N$ terms is written as follows:
\begin{center}
\begin{equation}
\ln K(\mathbf{x})\approx \left\langle \ln K(\mathbf{x}) \right\rangle +\sum\limits_{i=1}^{N}{\sqrt{{{\lambda }_{i}}}{{f}_{i}}(\mathbf{x}){{\xi }_{i}}}.
\label{eq:kle trunc}
\end{equation}
\end{center}
Once the covariance function of $\ln K(\mathbf{x})$ is specified, the eigenvalue and eigenfunction of its covariance can be determined, subsequently allowing parameterization of the stochastic field $\ln K(\mathbf{x})$ with a vector consisting of independent stochastic variables:
\begin{center}
\begin{equation}
\mathbf{\xi }=\left\{ {{\xi }_{1}},{{\xi }_{2}},\cdots ,{{\xi }_{N}} \right\}.
\end{equation}
\end{center}
Various samples of $\mathbf{\xi }$ can embody diverse realizations of $\ln K(\mathbf{x})$. Consequently, the estimation of hydraulic conductivity $K(\mathbf{x})$, denoted as $\hat{K}(\mathbf{x};\mathbf{\xi})$, undergoes updates by adjusting the vector $\mathbf{\xi}$ in the inversion process. Through this mechanism, the number of parameters requiring estimation in the inversion process can be reduced significantly, transitioning from a field to a vector. More detailed introductions about KLE can be found in previous relevant studies \citep{li2007probabilistic,zhang2004efficient}.

\subsection{PICD-based direct inversion}
\label{sec:PICD}
In the inversion process, many inversion methods typically necessitate repeated forward calculations to compute data mismatch and to evaluate the likelihood of different estimations, resulting in a substantial computational burden. In this work, we proposed a novel direct inversion framework employing the Physics-Informed Convolutional Decoder (PICD). In the PICD framework, the need for multiple numerical forward calculations is circumvented, enabling the direct estimation of unknown parameters using measurements and domain knowledge.

\begin{figure}[htb]
    \centering
    \includegraphics[width =\textwidth]{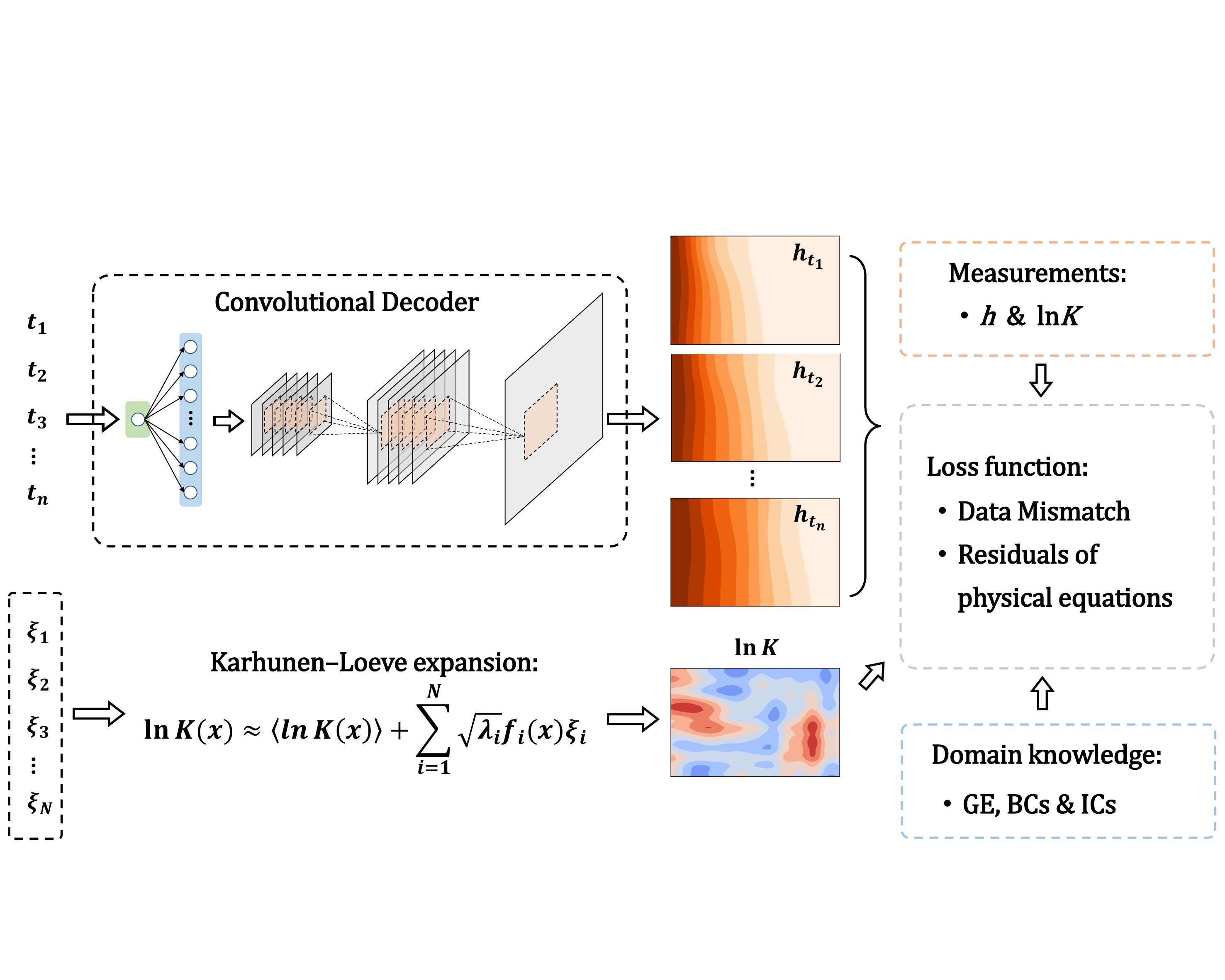}
    \caption{A schematic diagram of the proposed PICD framework. Here, GE stands for governing equation, BCs for boundary conditions, and ICs for initial conditions of the studied problem.}
    \label{fig:1}
\end{figure}

To approximate the dynamic and spatial distribution of hydraulic heads, $h(\mathbf{x},t)$, a convolutional decoder network is constructed, as shown in Figure \ref{fig:1}. In this network architecture, a single node is connected to the decoder as the input layer. Since convolutional networks excel at extracting spatial information, only the time information is required as input, resulting in a single node in the input layer. The decoder network receives time-step information as codes, producing decoded outputs that represent the spatial distribution of hydraulic heads, as illustrated in Figure \ref{fig:1}. The network can predict the evolution of hydraulic heads by inputting different time-steps, with the prediction denoted as $\mathbf{\hat{h}}(t;\gamma )$, where $\gamma$ represents the parameters of the convolutional decoder network. 

To align with the measurements of hydraulic heads, the training process aims to minimize the mismatch between the predictions and the observed data. The objective can be formulated as follows:
\begin{center}
\begin{equation}
{{L}_{h}}(\gamma )=\frac{1}{{{N}_{h}}}\sum\limits_{i=1}^{{{N}_{h}}}{{{\left| h({{\mathbf{x}}_{i}},{{t}_{i}})-\hat{h}({{\mathbf{x}}_{i}},{{t}_{i}};\gamma ) \right|}^{2}}},
\label{eq:mismatch}
\end{equation}
\end{center}
where $\hat{h}(\mathbf{x},t;\gamma )$ denotes the predicted hydraulic head at a specific coordinate $\mathbf{x}$ of $\mathbf{\hat{h}}(t;\gamma )$, and ${{N}_{h}}$ denotes the total number of available hydraulic head measurements. 

In the presence of hydraulic conductivity measurements, the objective is to minimize the mismatch between the estimations and measurements, as expressed as follows:
\begin{center}
\begin{equation}
{{L}_{K}}(\mathbf{\xi })=\frac{1}{{{N}_{K}}}\sum\limits_{i=1}^{{{N}_{K}}}{{{\left| \ln K({{\mathbf{x}}_{i}})-\ln \hat{K}({{\mathbf{x}}_{i}};\mathbf{\xi }) \right|}^{2}}},
\label{eq:mismatch_k}
\end{equation}
\end{center}
where ${{N}_{K}}$ denotes the total number of available hydraulic conductivity measurements. The parameterized vector, $\mathbf{\xi }$, is updated to refine the estimation of $K(\mathbf{x})$. Therefore, Eqs.~(\ref{eq:mismatch}) and~(\ref{eq:mismatch_k}) collectively constitute the data mismatch terms in the total loss function.

Incorporating domain knowledge into the inversion process is essential, beyond relying solely on available measurements. Here, domain knowledge encompasses the governing equation (GE), boundary conditions (BCs), and initial conditions (ICs) of the studied problem. Given that the convolutional decoder network offers grid-based or discretized predictions of hydraulic heads, the domain knowledge can also be transformed into discretized form to guide the training of the convolutional decoder network \citep{wang2021theory}. 

The governing equation of groundwater flow in Eq. (\ref{eq:ge}) can be discretized using the finite difference method (here, we focus on a two-dimensional (2D) problem):
\begin{center}
\begin{equation}
\begin{split}
  & {{S}_{s}}\frac{{{h}_{i,j,m}}-{{h}_{i,j,m-1}}}{\Delta t}-\frac{{{K}_{i+{1}/{2}\;,j}}\frac{{{h}_{i+1,j,m}}-{{h}_{i,j,m}}}{\Delta {{x}_{i+{1}/{2}\;}}}-{{K}_{i-{1}/{2,j}\;}}\frac{{{h}_{i,j,m}}-{{h}_{i-1,j,m}}}{\Delta {{x}_{i-{1}/{2}\;}}}}{\Delta {{x}_{i}}} \\ 
 & \text{  }-\frac{{{K}_{i,j+{1}/{2}\;}}\frac{{{h}_{i,j+1,m}}-{{h}_{i,j,m}}}{\Delta {{y}_{j+{1}/{2}\;}}}-{{K}_{i,j-{1}/{2}\;}}\frac{{{h}_{i,j,m}}-{{h}_{i,j-1,m}}}{\Delta {{y}_{j-{1}/{2}\;}}}}{\Delta {{y}_{j}}}-{{q}_{i,j,m}}=\text{0}, \\ 
 & \text{      (}i=1,\text{ }2,\text{ }...,\text{ }{{N}_{x}};\text{ }j=1,\text{ }2,\text{ }...,\text{ }{{N}_{y}};\text{ }m=1,\text{ }2,\text{ }...,\text{ }{{N}_{t}}\text{)}. \\ 
\end{split}
\end{equation}
\end{center}
Here $\Delta x$ and $\Delta y$ denote the interval size of the discretized grids in the horizontal and vertical directions, respectively. Additionally, $\Delta {{x}_{i+{1}/{2}\;}}$ denotes the distance of ${\left( \Delta {{x}_{i}}+\Delta {{x}_{i+1}} \right)}/{2}\;$, and a similar definition applies to $\Delta {{y}_{j+{1}/{2}\;}}$. The symbol $\Delta t$ denotes the discretization interval in time dimension. The subscript $i$, $j$ and $m$ denote the indexes of discretized grids in horizontal, vertical, and time dimensions, respectively. Furthermore, ${{N}_{x}}$, ${{N}_{y}}$, and ${{N}_{t}}$ denote the total number of discretized grids in horizontal, vertical, and time dimensions, respectively. The term ${{q}_{i,j,m}}$ denotes the sink/source term, which takes a value of zero at the grid without wells. The symbol ${{K}_{i+{1}/{2}\;,j}}$ denotes the hydraulic conductivity between grid $(i,\text{ }j)$ and $(i+1,\text{ }j)$, often represented as the mean of the hydraulic conductivity at the two adjacent grids. Specifically, the inter-grid conductivity is calculated with the harmonic mean of hydraulic conductivities at grid $(i,\text{ }j)$ and $(i+1,\text{ }j)$, as expressed follows:
\begin{center}
\begin{equation}
{{K}_{i+{1}/{2}\;,j}}=\frac{\Delta {{x}_{i+{1}/{2}\;}}}{\frac{\left( {1}/{2}\; \right)\Delta {{x}_{i}}}{{{K}_{i,j}}}+\frac{\left( {1}/{2}\; \right)\Delta {{x}_{i+1}}}{{{K}_{i+1,j}}}}.
\label{eq:hormonic}
\end{equation}
\end{center}
For a uniformly-discretized grid (i.e., $\Delta {{x}_{i}}=\Delta {{x}_{i+1}}$), Eq. (\ref{eq:hormonic}) can be simplified as ${{K}_{i+{1}/{2}\;,j}}=\frac{2{{K}_{i,j}}\cdot {{K}_{i+1,j}}}{{{K}_{i,j}}+{{K}_{i+1,j}}}$.

Utilizing the discretized form of the governing equation, we can guide the convolutional decoder network during the training process. This is crucial, as both the estimations of hydraulic conductivity and the predictions of hydraulic heads need to adhere to the governing equation. Subsequently, the approximated hydraulic head and conductivity are substituted into the discretized governing equation to calculate the residual, $R$. The residual at grid $(i,j)$ can assess how well the approximations align with the governing equation:
\begin{center}
\begin{equation}
\begin{split}
  & {{R}_{i,j}}(t;\mathbf{\xi },\gamma )={{S}_{s}}\frac{{{{\hat{h}}}_{i,j}}(t;\gamma )-{{{\hat{h}}}_{i,j}}(t-\Delta t;\gamma )}{\Delta t} \\ 
 & -\frac{{{{\hat{K}}}_{i+{1}/{2}\;,j}}(\mathbf{\xi })\frac{{{{\hat{h}}}_{i+1,j}}(t;\gamma )-{{{\hat{h}}}_{i,j}}(t;\gamma )}{\Delta {{x}_{i+{1}/{2}\;}}}-{{{\hat{K}}}_{i-{1}/{2,j}\;}}(\mathbf{\xi })\frac{{{{\hat{h}}}_{i,j}}(t;\gamma )-{{{\hat{h}}}_{i-1,j}}(t;\gamma )}{\Delta {{x}_{i-{1}/{2}\;}}}}{\Delta {{x}_{i}}} \\ 
 & -\frac{{{{\hat{K}}}_{i,j+{1}/{2}\;}}(\mathbf{\xi })\frac{{{{\hat{h}}}_{i,j+1}}(t;\gamma )-{{{\hat{h}}}_{i,j}}(t;\gamma )}{\Delta {{y}_{j+{1}/{2}\;}}}-{{{\hat{K}}}_{i,j-{1}/{2}\;}}(\mathbf{\xi })\frac{{{{\hat{h}}}_{i,j}}(t;\gamma )-{{{\hat{h}}}_{i,j-1}}(t;\gamma )}{\Delta {{y}_{j-{1}/{2}\;}}}}{\Delta {{y}_{j}}}-{{q}_{i,j}}(t), \\ 
 & \text{      (}i=1,\text{ }2,\text{ }...,\text{ }{{N}_{x}};\text{ }j=1,\text{ }2,\text{ }...,\text{ }{{N}_{y}}\text{)}. \\ 
 \label{eq:residual}
  \end{split}
\end{equation}
\end{center}
Here ${{\hat{h}}_{i,j}}(t;\gamma )$ denotes the predicted hydraulic head at grid $(i,j)$ and time-step, $t$. To force the estimations and predictions to adhere to the governing equation, the residual in Eq.~(\ref{eq:residual}) should be minimized during the training process. This constitutes the constraint of the governing equation in the loss function, as formulated below:
\begin{center}
\begin{equation}
{{L}_{GE}}(\mathbf{\xi },\gamma )=\frac{1}{{{N}_{x}}}\frac{1}{{{N}_{y}}}\frac{1}{{{N}_{t}}}\sum\limits_{i=1}^{{{N}_{x}}}{\sum\limits_{j=1}^{{{N}_{y}}}{\sum\limits_{m=1}^{{{N}_{t}}}{{{\left| {{R}_{i,j}}({{t}_{m}};\mathbf{\xi },\gamma ) \right|}^{2}}}}}.
\end{equation}
\end{center}

\begin{figure}[htb]
    \centering
    \includegraphics[width =5in]{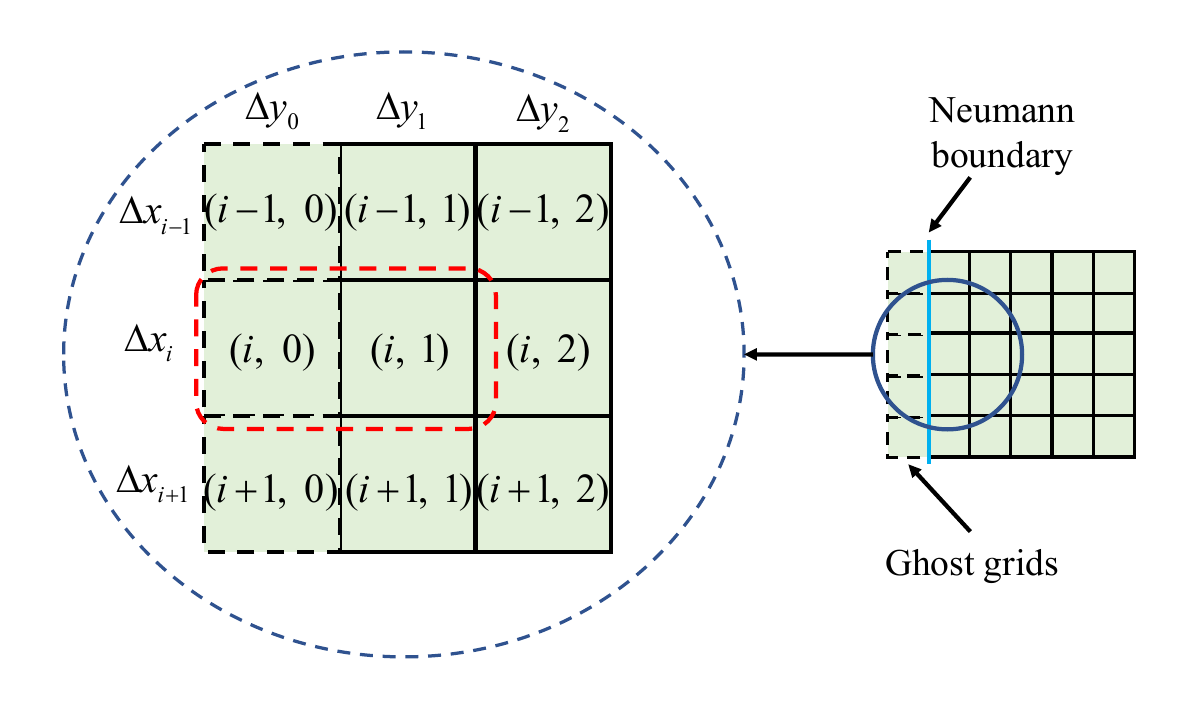}
    \caption{Diagram for illustration of ghost grids (boxes of black dashed line) for the Neumann boundary. Here, the red dashed box depicts how the flux in Eq.~\ref{eq:flux} is defined at the interface.}
    \label{fig:2}
\end{figure}

Similarly, the boundary and initial conditions can also be utilized as domain knowledge to guide the decoder network. The Dirichlet boundary condition in Eq.~(\ref{eq:dboun}) and the initial conditions in Eq.~(\ref{eq:ic}) are relatively straightforward to handle, and can be incorporated as constraints as follow:
\begin{center}
\begin{equation}
{{L}_{D}}(\gamma )=\frac{1}{{{N}_{D}}}\sum\limits_{i=1}^{{{N}_{D}}}{{{\left| \hat{h}({{\mathbf{x}}_{i}},{{t}_{i}};\gamma )-H({{\mathbf{x}}_{i}}) \right|}^{2}}},
\end{equation}
\end{center}
\begin{center}
\begin{equation}
{{L}_{I}}(\gamma )=\frac{1}{{{N}_{I}}}\sum\limits_{i=1}^{{{N}_{I}}}{{{\left| \hat{h}({{\mathbf{x}}_{i}},0;\gamma )-{{H}_{0}}({{\mathbf{x}}_{i}}) \right|}^{2}}}.
\end{equation}
\end{center}
Here, ${{N}_{D}}$ and ${{N}_{I}}$ denote the total number of grids used to impose the Dirichlet boundary and initial conditions, respectively. Concerning the Neumann boundary condition in Eq.~(\ref{eq:nboun}), the finite difference method can be applied for discretization. To represent the flux across the Neumann boundary, ‘ghost grids’ can be introduced along the boundary, as shown in Figure~\ref{fig:2}. These ghost grids can be included in the calculation but do not physically exist in the domain. As an illustration, for the interface between the grids $(i,0)$ and $(i,1)$ (enclosed in the red dashed box in Figure~\ref{fig:2}), the flux can be represented as:
\begin{center}
\begin{equation}
{{Q}_{i,{1}/{2}\;}}={{K}_{i,{1}/{2}\;}}\frac{{{h}_{i,1,m}}-{{h}_{i,0,m}}}{{\left( \Delta {{y}_{0}}+\Delta {{y}_{1}} \right)}/{2}\;},
\label{eq:flux}
\end{equation}
\end{center}
where subscript 0 represents the values of ghost grids. The flux is then integrated into the governing equation constructed for the boundary grids. As an illustration, for the boundary grid $(i,1)$, the discretized governing equation can be articulated by substituting Eq.~(\ref{eq:flux}):
\begin{center}
\begin{equation}
\begin{split}
\begin{aligned}
& S_s \frac{h_{i, 1, m}-h_{i, 1, m-1}}{\Delta t}-\frac{K_{i+1 / 2,1} \frac{h_{i+1,1, m}-h_{i, 1, m}}{\Delta x_{i+1 / 2}}-K_{i-1 / 2,1} \frac{h_{i, 1, m}-h_{i-1,1, m}}{\Delta x_{i-1 / 2}}}{\Delta x_i} \\
& -\frac{K_{i, 3 / 2} \frac{h_{i, 2, m}-h_{i, 1, m}}{\left(\Delta y_1+\Delta y_2\right) / 2}-Q_{i, 1 / 2}}{\Delta y_1}-q_{i, 1, m}=0 \;\;\;.
\label{eq:nboun_inserted_ge}
\end{aligned}
\end{split}
\end{equation}
\end{center}

Utilizing Eq.~(\ref{eq:nboun_inserted_ge}), the residuals ${{R}_{N}}(\mathbf{x},t;\mathbf{\xi },\gamma )$ can be calculated for the grids on the Neumann boundary. Consequently, the constraints of the Neumann boundary can be imposed by incorporating the conditions into the governing equation for the boundary grids, and the loss function can be expressed as:
\begin{center}
\begin{equation}
{{L}_{N}}(\mathbf{\xi },\gamma )=\frac{1}{{{N}_{N}}}\frac{1}{{{N}_{t}}}\sum\limits_{i=1}^{{{N}_{N}}}{\sum\limits_{m=1}^{{{N}_{t}}}{{{\left| {{R}_{N}}({{\mathbf{x}}_{i}},{{t}_{m}};\mathbf{\xi },\gamma ) \right|}^{2}}}},
\end{equation}
\end{center}
where ${{N}_{N}}$ denotes the total number of grids on the Neumann boundary. 

Finally, the total loss function for the physics-informed convolutional decoder framework can be formulated as:
\begin{center}
\begin{equation}
\begin{split}
L(\mathbf{\xi },\gamma )={{\lambda }_{K}}{{L}_{K}}(\mathbf{\xi })+{{\lambda }_{h}}{{L}_{h}}(\gamma )+{{\lambda }_{GE}}{{L}_{GE}}(\mathbf{\xi },\gamma ) \\ 
+{{\lambda }_{D}}{{L}_{D}}(\gamma )+{{\lambda }_{I}}{{L}_{I}}(\gamma )+{{\lambda }_{N}}{{L}_{N}}(\mathbf{\xi },\gamma ), \\ 
\label{eq:loss func}
\end{split}
\end{equation}
\end{center}
where $\lambda$ denotes the weight assigned to different terms in the loss function. It is evident that the parameterized hydraulic conductivity field with vector $\mathbf{\xi }$ undergoes tunning, akin to the adjustment of network parameters during the training process. Upon completion of training, the trained network can predict the distribution of hydraulic heads at different times, while the tuned vector $\mathbf{\xi }$ can provide the estimated conductivity field by inputting it into the expansion form of KLE. Throughout the training process, both the available measurements and domain knowledge are leveraged to refine the estimations. Notably, in this inversion process, repeated forward calculations are unnecessary, and the mathematical model of forward modeling is directly employed to constrain the decoder network -- an approach distinct from common inversion methods.

\section{Results}
\label{sec:results}
In this section, numerical experiments of groundwater flow problems are conducted to assess the performance of the proposed physics-informed convolutional decoder (PICD) direct inversion framework. A two-dimensional (2D) groundwater flow case is considered, where the physical domain is a rectangle with a length of ${{L}_{x}}=1200\text{ m}$ and a width of  ${{L}_{y}}=640\text{ m}$. The left and right boundaries are constant head boundaries with specified heads of ${{h}_{x=0}}=305\text{ m}$ and ${{h}_{x={{L}_{x}}}}=300\text{ m}$, respectively. The upper and lower boundaries are no-flow boundaries. The initial state of the hydraulic head is $300\text{ m}$ uniformly in the domain, except at the left boundary. 

The log-transformed hydraulic conductivity field is assumed to follow a normal distribution with a mean of $\left\langle \ln K \right\rangle =1$ and a standard deviation of ${{\sigma }_{\ln K}}=1$. The covariance function of the stochastic field, $\ln K$, is specified to be of a separable exponential type:
\begin{center}
\begin{equation}
{{C}_{\ln K}}(\mathbf{x},\mathbf{x'})=\sigma _{\ln K}^{2}\exp \left( -\frac{\left| x-x' \right|}{{{\eta }_{x}}}-\frac{\left| y-y' \right|}{{{\eta }_{y}}} \right),
\end{equation}
\end{center}
where $\mathbf{x}=(x,y)$ and $\mathbf{x'}=(x',y')$ denote two points in the studied physical domain, ${{\eta }_{x}}$ and ${{\eta }_{y}}$ denote the correlation lengths of the stochastic field in the $x$ and $y$ directions, respectively. In this case, the correlation lengths are set to ${{\eta }_{x}}=0.4{{L}_{x}}=480\text{ m}$ and ${{\eta }_{y}}=0.4{{L}_{y}}=256\text{ m}$. 

Given a specific hydraulic conductivity field, this groundwater flow case can be simulated using the MODFLOW numerical simulator \citep{harbaugh2005modflow} to obtain the corresponding hydraulic head distribution. In this work, the physical domain is discretized into $60\times 32$ grid blocks with a grid size of $\Delta x=\Delta y=20\text{ m}$. The simulation is computed using a time-step of $\Delta t=0.2\text{ day}$ and terminated at $t =10$~days. Hydraulic heads and conductivities of specific (monitoring) locations are utilized as known measurements during the inversion process for the estimation of the distribution of the hydraulic conductivity field of the whole simulated domain. 

\subsection{Inversion of hydraulic conductivity generated with KLE}
\label{sec:kle generated case}

\subsubsection{Base case}
\label{sec:base case}
\begin{figure}[htbp]
    \centering
    \includegraphics[width =\textwidth]{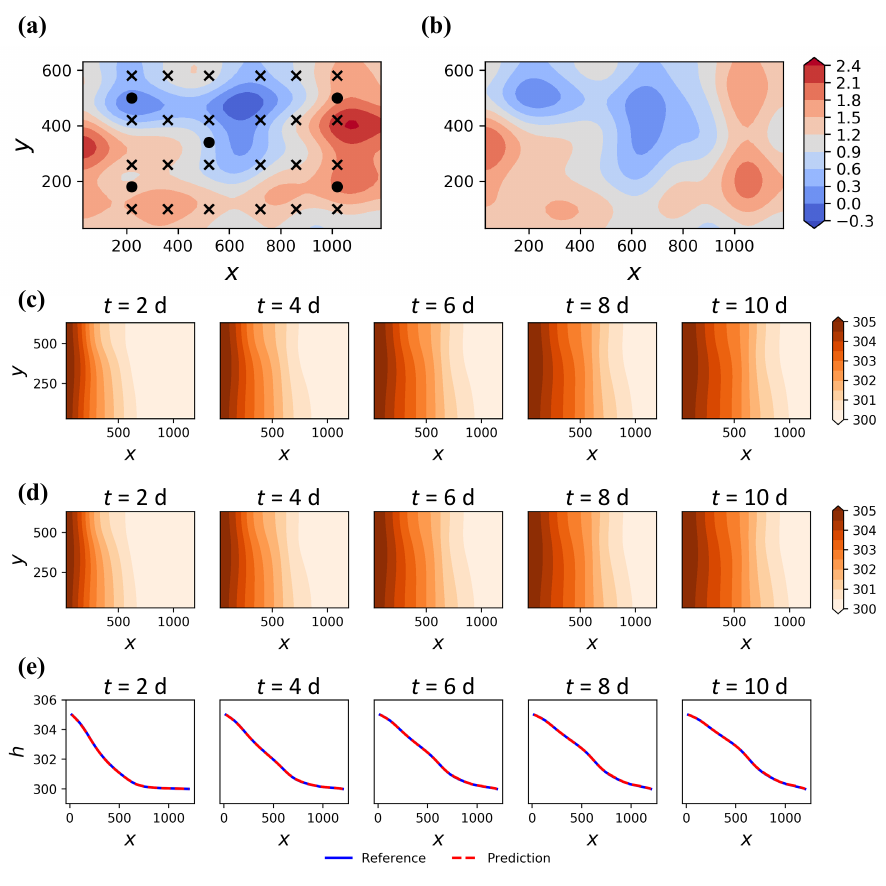}
    \caption{Inversion of $\ln K$ for the base case: (a) Reference field of $\ln K$ and monitoring locations (the cross symbol for hydraulic head and the dot symbol for hydraulic conductivity); (b) Estimation of $\ln K$ using the proposed PICD framework; hydraulic head comparison for the base case: (c) The reference hydraulic head distribution; (d) The predicted hydraulic head distribution; (e) The slicing comparison of hydraulic head at $y=320\text{ m}$.}
    \label{fig:3}
\end{figure}

In this subsection, the base case is presented -- a relatively simple scenario in which the reference conductivity field is randomly generated using Karhunen–Loeve expansion (KLE). Since KLE is also employed to approximate the conductivity field in our framework, the KLE-generated field is chosen to be easier for the estimation as a comparable field. The reference field of $\ln K$ is shown in Figure~\ref{fig:3}(a). During the field generation using KLE, 85\% of the energy of the stochastic field is preserved, leading to 32 retained terms in the expansion. Therefore, the dimension of vector $\mathbf{\xi }$ is also 32. 

In this base case, 24 monitoring locations are placed in the domain to collect time-series of hydraulic heads, as illustrated using the cross symbols in Figure~\ref{fig:3}(a). Hydraulic conductivity is assumed to be available at five observation points, which are denoted with black dots in Figure~\ref{fig:3}(a). These monitored hydraulic heads and five available hydraulic conductivities are used as available measurements to perform the inversion of the unknown conductivity field.

In this base case, a convolutional decoder is constructed to approximate the hydraulic head distribution, and the details of the network architecture are presented in Table~\ref{table:1}. The log hydraulic conductivity field is approximated with the KLE expression shown in Eq.~(\ref{eq:kle trunc}). In the proposed PICD framework, the constructed network and the vector $\mathbf{\xi }$ in KLE are then trained with the available measurements and domain knowledge. The learning rate is set to be 0.001, and the total iteration number is set to be 20,000. After training, the decoder can predict the evolution and distribution of hydraulic heads, and the estimated hydraulic conductivity can be obtained by inputting the optimized vector $\mathbf{\xi }$ into the KLE expression. The estimated hydraulic conductivity field is presented in Figure~\ref{fig:3}(b), demonstrating the accuracy of the PICD framework in reproducing the hydraulic conductivity closely resembling the reference field. Additionally, the comparison between the predicted and reference hydraulic heads over time is also presented in Figure~\ref{fig:3}. The reference hydraulic heads are presented in Figure~\ref{fig:3}(c), and the predicted hydraulic heads are presented in Figure~\ref{fig:3}(d). Furthermore, the comparison of hydraulic head slicing at $y=320\text{ m}$ is presented in Figure~\ref{fig:3}(e). The results demonstrate that the predictions closely match the references, showcasing the PICD's ability to achieve satisfactory forward modeling performance in the inversion process. 

\begin{figure}[htb]
    \centering
    \includegraphics[width =\textwidth]{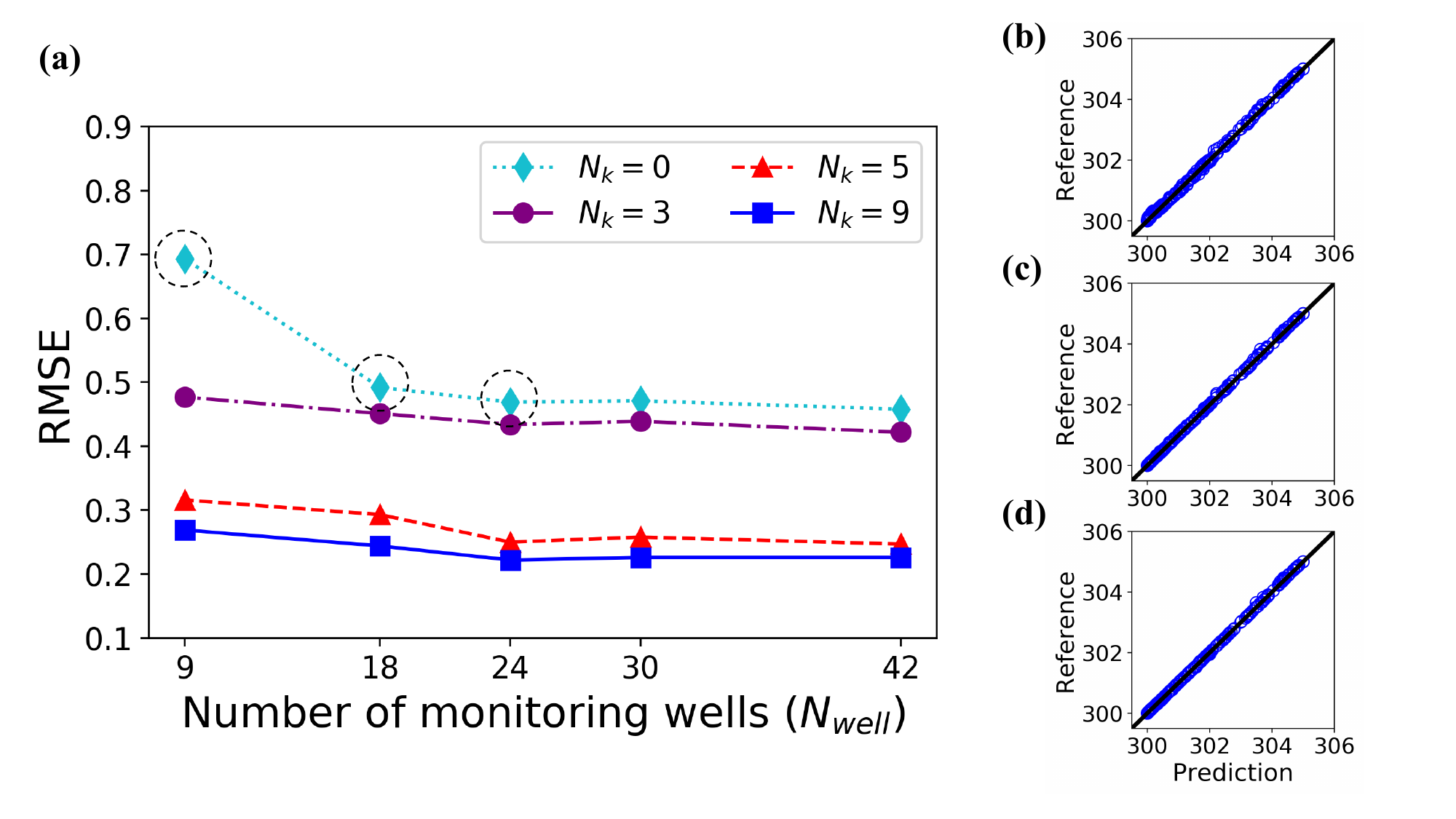}
    \caption{(a) The Root Mean Square Error (RMSE) of inversion results with varying numbers of available measurements; (b) Correlation plot of predicted and reference hydraulic heads when ${{N}_{k}}=0$ and ${{N}_{well}}=9$; (c) Correlation plot of predicted and reference hydraulic heads when ${{N}_{k}}=0$ and ${{N}_{well}}=18$; (d) Correlation plot of predicted and reference hydraulic heads when ${{N}_{k}}=0$ and ${{N}_{well}}=24$.}
    \label{fig:4}
\end{figure}

\begin{table}[htbp]
\centering
\caption{The architecture details of the utilized convolutional decoder network for the base case.}
\makebox[\textwidth][c]{
\begin{tabular}{ccc}\hline
\pmb{Layers}&\pmb{Output size}&\pmb{Number of channels}\\\hline
Fully-connected layer (input size 1)&	100&	/\\
Fully-connected layer (input size 100)&	2688&	/\\
Reshape (input size 2688)&	3×7	&128\\
Deconvolution layer (kernel 4×3, stride 2, padding 0)&	8×15	&64\\
Activation (Swish)&	8×15&	64\\
Deconvolution layer (kernel 3×3, stride 2, padding 1)&	15×29&	32\\
Activation (Swish)&	15×29&	32\\
Deconvolution layer (kernel 4×4, stride 2, padding 1)&	30×58	&16\\
Activation (Swish)&	30×58	&16\\
Deconvolution layer (kernel 3×3, stride 1, padding 0)&	32×60&	1\\
Activation (Sigmoid)&	32×60&	1\\\hline
\end{tabular}
}
\label{table:1}
\end{table}

Next, we examine the effect of the amount available measurements (hydraulic heads and conductivity as indirect and direct measurements, respectively) by performing the inversion with different numbers of monitoring locations (${{N}_{well}}$) for hydraulic head measurements and different numbers of available conductivity measurements (${{N}_{k}}$). The inversion results with different available measurements are presented in Figure~\ref{fig:4}(a). It is evident that the inversion results become increasingly accurate as the number of available measurements increases, whether direct or indirect measurements, aligning with common sense. Notably, even with limited direct measurements (i.e., conductivity), the inversion results remain satisfactory when sufficient hydraulic heads are available. In an extreme case with no conductivity measurements, the proposed PICD framework can still provide relatively accurate estimations of the reference field based solely on hydraulic heads, yielding accurate prediction of hydraulic head distribution. Three challenging situations, marked with dashed circles in Figure~\ref{fig:4}(a), represent the least accurate inversion results in our settings. The corresponding predictions of hydraulic heads in these situations are sampled and compared with the reference heads, as shown in Figures~\ref{fig:4}(b), (c), and (d). Despite imperfect estimations of the conductivity field, the predicted hydraulic heads still closely match the reference heads, demonstrating the PICD's ability to achieve satisfactory forward modeling even with very limited available measurements. 

\begin{figure}[htbp]
    \centering
    \includegraphics[width =\textwidth]{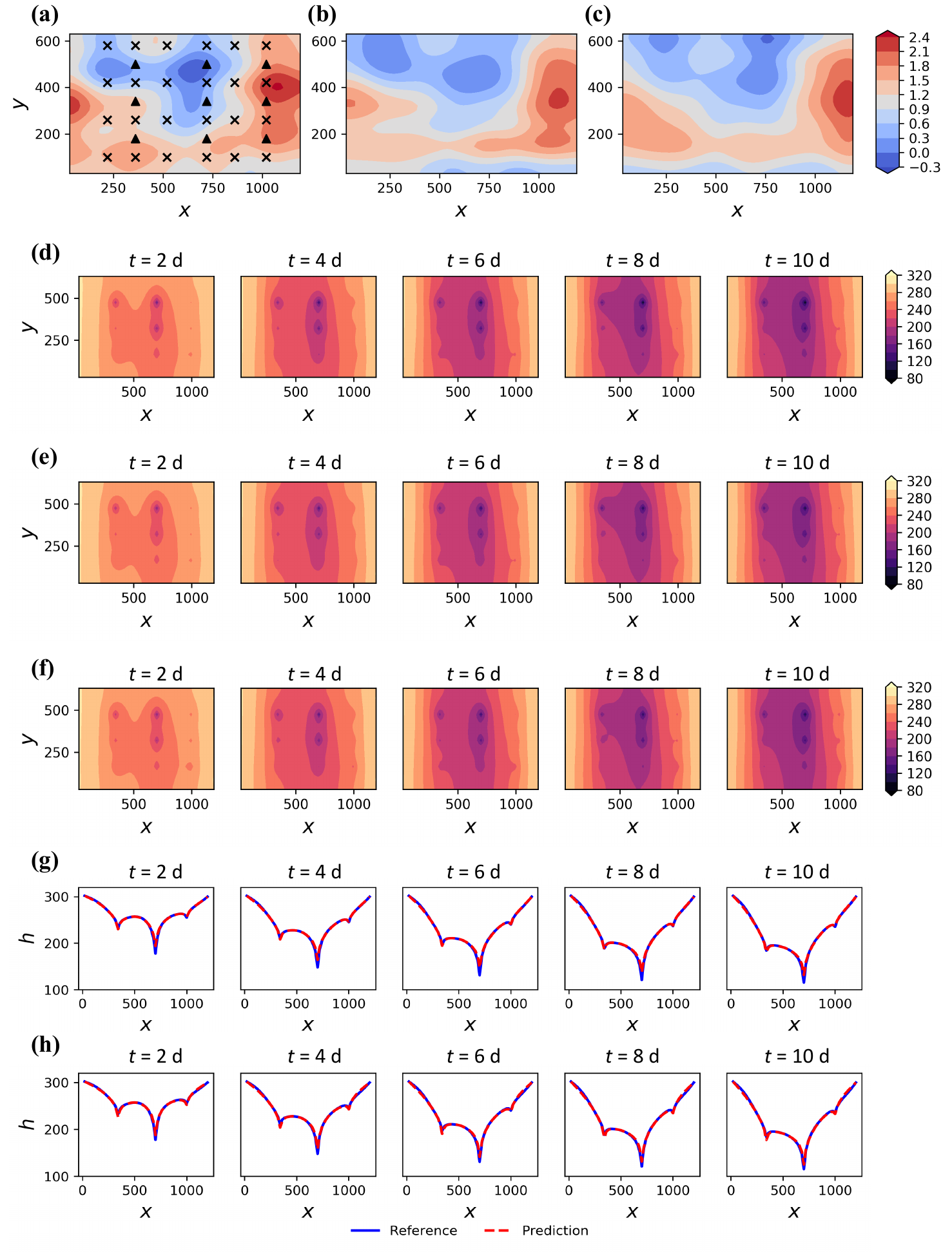}
    \caption{Inversion of $\ln K$ for the case with pumping wells: (a) Reference field of $\ln K$ and well locations (note that the black triangle symbol indicates the pumping well, and the cross symbol indicates the monitoring location for the hydraulic head); (b) Estimation of $\ln K$ using both hydraulic head and conductivity; (c) Estimation of $\ln K$ using only hydraulic head; hydraulic head estimation during the inversion of $\ln K$ for the case with pumping wells: (d) References of hydraulic head distribution; (e) Predictions of hydraulic head distribution using both hydraulic head and conductivity measurements; (f) Predictions of hydraulic head distribution using only hydraulic head measurements; (g) Comparison of hydraulic head of slice at $y=320\text{ m}$ for (e); (h) Comparison of hydraulic head of slice at $y=320\text{ m}$ for (f). Here the time is presented as day (d).}
    \label{fig:5}
\end{figure}

\subsubsection{Case with pumping wells}
\label{sec:pumping well case}
In this subsection, we investigate a more challenging scenario by incorporating nine pumping wells as sink/source terms, whose dynamics are more complex to capture. The locations of these wells, denoted with black triangles in Figure~\ref{fig:5}(a), pump water at a rate of 200 ${{\text{m}}^{\text{3}}}\text{/day}$. The hydraulic conductivities at the pumping well locations are assumed to be available. Additionally, hydraulic heads are collected over time in 24 monitoring wells, whose locations are the same as those in the base case in Section~\ref{sec:base case}, as shown in Figure~\ref{fig:5}(a). The collected measurements can then be utilized to perform the inversion of the unknown conductivity field. The network architecture and other parameter settings are consistent with those in the base case. Through simultaneous training of the network and the vector of random variable $\mathbf{\xi }$, the estimation of hydraulic head and conductivity can be obtained upon completion of the training process. 

The estimated conductivity field is presented in Figure~\ref{fig:5}(b), demonstrating its resemblance to the reference field. Both the high-conductivity and low-conductivity regions are well captured, underscoring the effectiveness and accuracy of inversion. Subsequently, we evaluate the accuracy of the approximated hydraulic head distribution over time with the network. The predicted hydraulic heads are presented in Figure~\ref{fig:5}(e), while the ground truth is given in Figure~\ref{fig:5}(d). In Figure~\ref{fig:5}(g), a comparison of slicing at $y=320\text{ m}$ between the prediction and ground truth is presented. Despite some spikes at the pumping wells, the network effectively captures the discontinuity in the hydraulic head distribution, demonstrating that the distribution of hydraulic head in the domain can be learned from the measurements and domain knowledge in the inversion process. 

Inversions with and without direct measurements of hydraulic conductivity in the case with pumping wells are also compared. In the scenario with direct measurements of $K$, hydraulic conductivity at pumping wells is assumed to be measured and utilized in the inversion process. Conversely, in the scenario without direct measurements of $K$, we perform the inversion solely with available hydraulic heads, omitting the ${{\lambda }_{K}}{{L}_{K}}(\mathbf{\xi })$ term in the loss function (Eq.~\ref{eq:loss func}). The estimated conductivity is presented in Figure~\ref{fig:5}(c). The estimation in the scenario without available $K$ is understandably less accurate than the outcome obtained with conductivity observations shown in Figure~\ref{fig:5}(b), due to the utilization of less information in the inversion process. However, it is worth noting that the general pattern of the reference field can still be captured even without direct conductivity measurements. Besides, the prediction and comparison of hydraulic heads are presented in Figure~\ref{fig:5}(f) and (h). It can be seen that the predictions of hydraulic head without $K$ measurements are consistent with the references. This implies that the performance of the proposed PICD framework exhibits reasonable robustness in scenarios lacking direct observations of the inverted quantities. 

\subsection{Inversion of hydraulic conductivity generated with SGSIM}
\label{sec:sgsim generated case}
\begin{figure}[htbp]
    \centering
    \includegraphics[width =\textwidth]{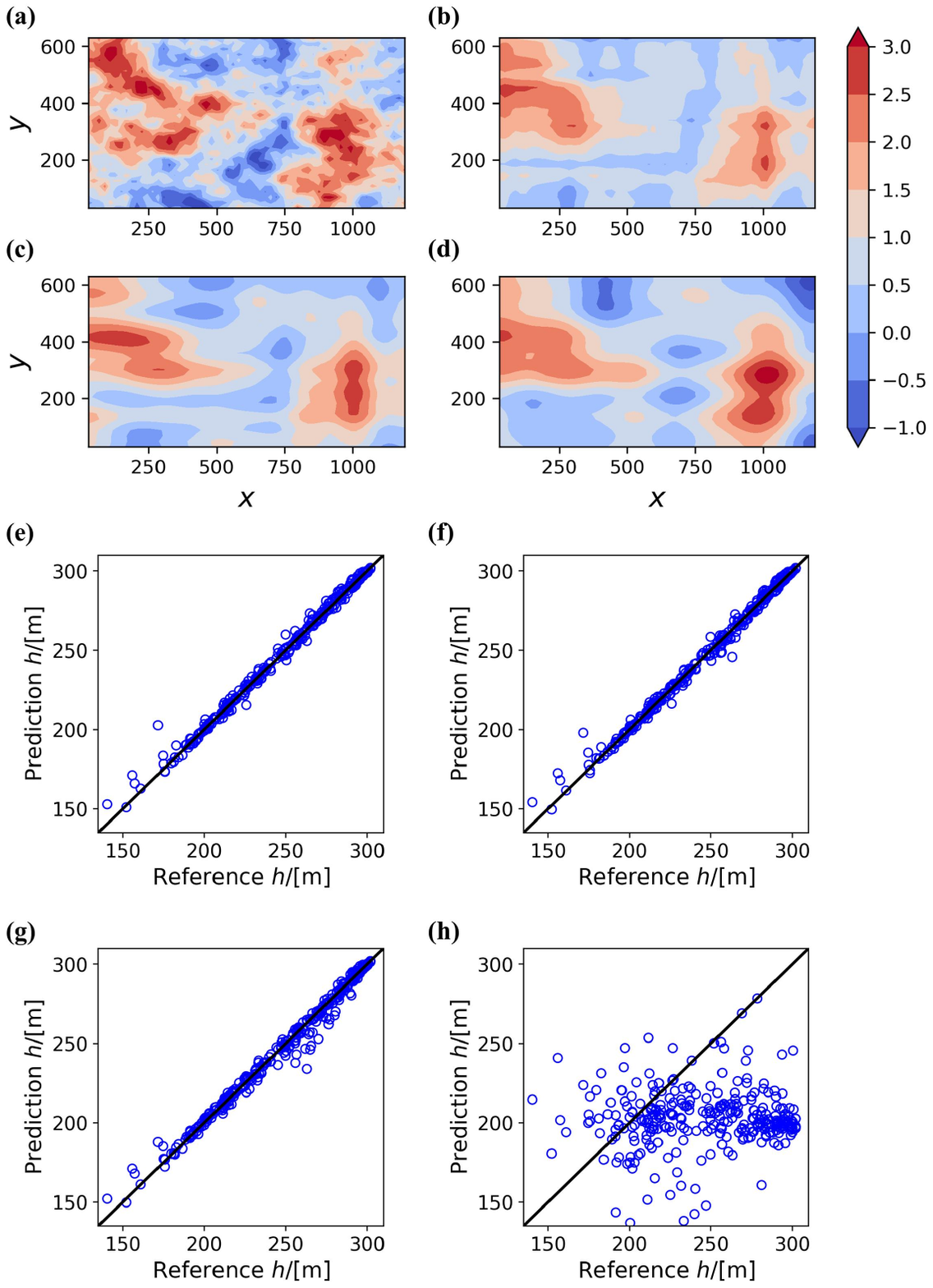}
    \caption{ (a) Reference field of $\ln K$ generated with SGSIM; (b) Estimation of $\ln K$ with KLE preserving 95\% energy; (c) Estimation of $\ln K$ with KLE preserving 90\% energy; (d) Estimation of $\ln K$ with KLE preserving 85\% energy; correlation plots of predicted and reference hydraulic heads: (e) KLE preserving 95\% energy; (f) KLE preserving 90\% energy; (g) KLE preserving 85\% energy; and (h) without domain knowledge.}
    \label{fig:6}
\end{figure}

In previous scenarios, the reference $K$ fields were generated using the KLE method, typically resulting in relatively smooth fields that theoretically facilitate the inversion tasks. In this subsection, we consider a more complex scenario in which the unknown reference $K$ field is generated using the Sequential Gaussian Simulation (SGSIM) method \citep{nussbaumer2018accelerating}, relying on statistical information. The generated reference field of log hydraulic conductivity,$\ln K$, is illustrated in Figure~\ref{fig:6}(a), and it can be seen that the field is less smooth than those in the former cases. The locations of pumping and monitoring wells remain consistent with the case in Section~\ref{sec:pumping well case}. 

In the inversion process, KLE is still employed to approximate hydraulic conductivity, and different percentages of energy are preserved while formulating the KLE to achieve varying fidelity of estimations. Specifically, we preserve 95\%, 90\%, and 85\% of the energy of the stochastic field while truncating the expansion, resulting in 164, 60, and 32 retained terms in KLE, respectively. Therefore, in these three different settings, the dimensions of the stochastic vector $\mathbf{\xi }$ to be trained are 164, 60, and 32, respectively. The performance of the proposed PICD framework is then assessed for these three scenarios, maintaining other settings the same as those in the previous case in Section~\ref{sec:pumping well case}. The corresponding inversion results are presented in Figures~\ref{fig:6}(b), (c), and (d), respectively. 

Observing these results, we note that the estimations exhibit different smoothness with varying percentages of energy preserved. However, all estimations demonstrate a similar distribution to the reference field, emphasizing that the general pattern of the conductivity field can be captured by the KLE representation even at different fidelity levels. Additionally, correlation plots between the predicted hydraulic heads from the three scenarios and the reference heads are presented in Figures~\ref{fig:6}(e), (f), and (g). It is evident that, with different retained energy levels of KLE, the predictions of hydraulic head remain satisfactory, even under low-fidelity (i.e., low retained energy) settings of KLE. This suggests that a rough estimation of hydraulic conductivity can be utilized to predict hydraulic heads with the incorporation of domain knowledge.

To further explore the effectiveness of domain knowledge, we use the decoder network to fit the available hydraulic heads, and subsequently predict the hydraulic head distribution without involving domain knowledge. The correlation plot between the predictions and reference values is presented in Figure~\ref{fig:6}(h). It is evident that the predictions without the domain knowledge differ significantly from the reference ones, exhibiting a rather poor correlation. This emphasizes the substantial benefits of incorporating domain knowledge in enhancing the predictive capabilities of the proposed PICD framework.

\subsection{Inversion of hydraulic conductivity with unknown correlation length}
\label{sec:unknown correlation length}

\begin{figure}[htb]
    \centering
    \includegraphics[width =\textwidth]{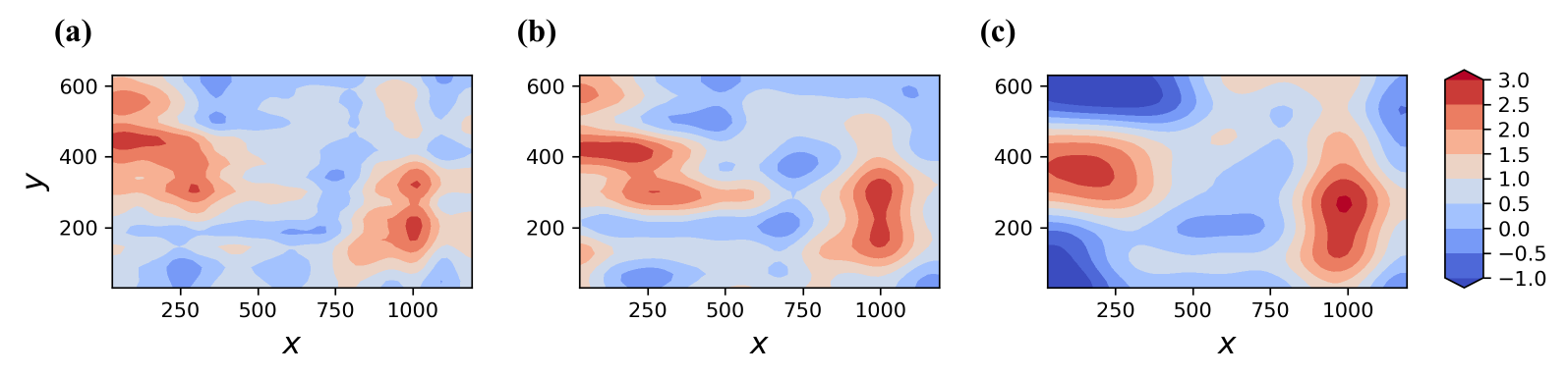}
    \caption{Estimation of $\ln K$ with inaccurate correlation lengths: (a) ${{\eta }_{x}}=240$ m, ${{\eta }_{y}}=128$ m; (b) ${{\eta }_{x}}=360$ m, ${{\eta }_{y}}=192$ m; and (c) ${{\eta }_{x}}=600$ m, ${{\eta }_{y}}=320$ m. }
    \label{fig:10}
\end{figure}

In previous cases, the correlation length of the hydraulic conductivity field is assumed to be known and is utilized in the KLE. To evaluate the performance of the proposed PICD framework under conditions with unknown correlation length, we intentionally use inaccurate values of the correlation length in the KLE during the inversion process. In this task, the reference hydraulic conductivity field is generated with SGSIM, which is the same as in Section~\ref{sec:sgsim generated case}. In the inversion, three inaccurate correlation lengths are used: 1) ${{\eta }_{x}}=240$ m, ${{\eta }_{y}}=128$ m; 2) ${{\eta }_{x}}=360$ m, ${{\eta }_{y}}=192$ m; and 3) ${{\eta }_{x}}=600$ m, ${{\eta }_{y}}=320$ m. While 85\% energy of the stochastic field is preserved to truncate the expansion in the KLE, other settings remain the same as those in previous cases. The inversion results are presented in Figure~\ref{fig:10}. It is observed that the correlation length settings influence the local heterogeneity and smoothness of the estimated fields. Nonetheless, from a global perspective, all estimations display a similar distribution to the reference field, particularly in terms of the location of highly permeable regions. In essence, even when the correlation length of the reference field is unknown or inaccurately assumed, the proposed PICD framework can still provide estimations that relatively accurately reflect the general distribution of the unknown conductivity field. This indicates that the PICD framework's requirement for precise prior information is relatively low. 

\section{Discussion and conclusions}
\label{sec:discussions and conclusions}
In this work, we introduce a novel direct inversion framework for groundwater flow problems, leveraging  deep learning techniques to harness domain knowledge and streamline the inversion process. The proposed Physics-Informed Convolutional Decoder (PICD) framework combines a convolutional network for approximating the hydraulic head distribution with Karhunen-Loeve expansion (KLE) for representing the hydraulic conductivity distribution. Both components are fined-tuned simultaneously during training to match available measurements and minimize discretized equation residuals, incorporating the governing equation, boundary conditions, and initial conditions of groundwater flow problems as domain knowledge. We use the finite difference method for the discretization of equations. After training, the decoder network predicts hydraulic heads, and when substituting the adjusted stochastic vectors into the KLE yields the estimated conductivity field. Therefore, in this inversion process, the repeated forward calculations of the physical model are circumvented, and the physical model is utilized as domain knowledge instead.  

The proposed PICD framework was rigorously evaluated across various groundwater flow scenarios, demonstrating its ability to satisfactorily estimate unknown conductivity fields even with sparse measurements. The predictions of hydraulic heads distribution are also satisfactory, showcasing robust performance across cases with or without pumping wells. Moreover, when dealing with conductivity fields generated using Sequential Gaussian Simulation (SGSIM), the proposed framework exhibited the capacity to provide relatively accurate estimations of the general distribution pattern. By adjusting the percentage of preserved energy in the KLE, estimations with varying fidelity were obtained. Additionally, in scenarios with unknown or inaccurately specified prior information, such as the correlation length of the conductivity fields, the framework still provided useful estimations of the fields, identifying highly permeable regions. This highlights the framework's potential to alleviate reliance on precise prior information during inversion.  

While the proposed PICD framework presents a promising approach by integrating both available measurements and domain knowledge without the need for extensive forward calculations, it has certain limitations. Notably, it provides deterministic estimations and lacks uncertainty quantification for inversion results. Addressing this limitation could involve incorporating Bayesian inference into the training process, representing a potential avenue for future research. The proposed framework offers a valuable contribution by redefining the inversion paradigm, emphasizing the simultaneous utilization of measurement data and domain knowledge for enhanced efficiency and accuracy in groundwater flow studies.

\section*{Acknowledgements}
This work was supported and partially funded by the National Natural Science Foundation of China (Grant No. 52288101) and the National Center for Applied Mathematics Shenzhen (NCAMS).

\bibliographystyle{elsarticle-harv} 
\bibliography{0Ref}

\end{document}